\documentclass[twocolumn,aps,prl,showpacs,amsmath,floatfix]{revtex4}
\usepackage{graphicx,bm,color}

\begin{document}
\title{Exotic Grazing Resonances in Nanowires}
\author{Simin Feng}
\author{Klaus Halterman}
\affiliation{Research and Intelligence Department \\  Naval Air Warfare Center, China Lake, CA 93555}
\date{\today}

\begin{abstract}
We investigate electromagnetic scattering from nanoscale wires and reveal
for the first time, the emergence of a family of exotic resonances, or
enhanced fields, for source waves close to grazing incidence. These grazing
resonances can have a much higher Q factor, broader bandwidth, and are much
less susceptible to material losses than the well known surface plasmon
resonances found in metal nanowires. Contrary to surface plasmon resonances
however, these grazing resonances can be excited in both dielectric and
metallic nanowires and are insensitive to the polarization state of the
incident wave. This peculiar resonance effect originates from the excitation
of long range guided surface waves through the interplay of coherently
scattered continuum modes coupled with the azimuthal first order propagating
mode of the cylindrical nanowire. The nanowire resonance phenomenon revealed
here can be utilized in broad scientific areas, including: metamaterial
designs, nanophotonic integration, nanoantennas, and nanosensors.

\end{abstract}


\maketitle
With the ever increasing advances in nanofabrication techniques, many
remarkable optical phenomena have been discovered, thus stimulating
considerable interest in light scattering at the nanometer length scale.
Within the emerging field of nanoplasmonics, metallic nanostructures can be
tailored to harness collective optical effects, namely, surface plasmon resonances (SPRs).
Localized field enhancement is one of the key underlying physical
characteristics in promising new nanotechnological applications.  Due to
relatively high energy confinement [Q factor] and small mode
volume, SPRs have been used in the construction of tightly packed photonic
devices\cite{Maier}, and play a major role in such applications\cite{Nie,Xu,Catchpole,Rockstuhl,Halterman,Feng,Valentine,Shalaev} as, surface-enhanced
Raman scattering\cite{Nie,Xu}, photoconversion\cite{Catchpole,Rockstuhl}, and Metamaterial designs\cite{Valentine,Shalaev}.  However, SPRs
are extremely sensitive to material loss, which diminishes the Q factor
significantly and can limit its benefits for the envisaged applications.
Discovery of another subwavelength field-enhancement mechanism would be
greatly beneficial in future nanoscale research.

It is well known that surface plasmons cannot be excited by TM polarization ($\bm{H}$ is perpendicular to the axis of the rod)\cite{Kottmann,Luk}.  With TE polarization SPRs can be strongly excited at normal incidence, but diminish as the incident angle approaches grazing incidence.  Grazing scattering has a broad applicability\cite{Rousseau,Gu,Lind,Barrick} and is a nontrivial and longstanding problem due to singularities arising at the zero grazing angle.  As the grazing angle approaches zero, intrinsic singularities in the scattering solutions reveal possible numerical instabilities.  The general solution intended for non-grazing incidence often fail to provide a clear picture of this limiting case due to divergences arising at very small grazing angles.  Even in the macroscopic and long wavelength regime, descriptions of grazing behaviors often leads to contradictions\cite{Barrick}.  At the grazing, scattering and propagation are inextricably linked via surface waves.  Physically meaningful concepts like scattering cross section and efficiency $Q$ must therefore be redefined.

In this Letter, we investigate the unexplored and nontrivial grazing
phenomena in long nanowires.  We undertake the analytic and numerical
challenges inherent to grazing scattering at the nanoscale and develop a
robust analytic solution to Maxwell's equations for grazing incidence.
Proper account of the guided surface waves is accomplished by generalizing
the definition of scattering efficiency.  Our calculations reveal a family of
non-plasmonic resonances with larger Q factors, broader bandwidth and are
more sustainable to material loss than SPRs.  It is determined that these
resonances originate from the excitation of guided asymmetric surface waves
corresponding to the $n=1$ mode of the nanorods, akin to first order
Sommerfeld waves, traditionally ascribed to the symmetric $(n=0)$ case for TM
waves propagating along a conducting cylinder.  These cyclic higher order $(n\neq0)$ periodic solutions have been historically overlooked, likely due to
the typically high attenuation at long wavelengths.  It is remarkable that
indeed grazing light can intricately couple to these first order cyclic
surface waves and contribute to exotic grazing resonances.  Contrary to the
quasistatic nature\cite{Fredkin} of SPRs, these grazing resonances are strictly
non-electrostatic.  This phenomenon, i.e. non-electrostatic resonances in the subwavelength 
regime, was brought into attention in recent work involving the effect of volume plasmons on 
enhanced fields in ultrasmall structures\cite{Feigenbaum}, revealing the richness of resonant 
scatterings in the nanoscale.

Consider an infinitely long cylinder of radius $a$ with its rotation axis along the $z$-direction [cylindrical coordinates
$(\rho,\phi,z)$], and with permittivity $\epsilon_1$ and permeability $\mu_1$.  
The background permittivity $\epsilon_0$ and permeability $\mu_0$, correspond to that in vacuum.  A plane TM wave is incident onto the cylinder with a grazing angle $\theta$ with respect to the z-axis.  Without loss of generality, the incident wave resides in the x-z plane with the harmonic time dependence, $\exp(-i\omega t)$.  
TM and TE modes are coupled in the scattered field due to the non-normal incidence.  The general solution\cite{Bohren} generally involves an infinite series of Bessel functions, and
often poses numerical difficulties for small grazing angles.  We take an alternative approach to derive the general solution.  The starting point is the vector Helmholtz equation in cylindrical coordinates, which is first solved for $E_z$ and $H_z$.  The remaining transverse field components are then easily determined\cite{Jackson}.
By matching tangential components of the fields at the cylinder surface $(\rho=a)$, the general solution for arbitrary incident angles can be obtained.  The grazing solution is then derived by asymptotic expansion of the general solution at the singularity $\theta=0$.  This approach has the advantage of a simpler analytical solution (written solely in terms of elementary functions), and possessing only the 0 and 1 azimuthal modes.  The detailed derivation of the general and grazing solutions are beyond the scope of this letter, and will be presented elsewhere\cite{Feng2}.  Here it suffices to provide the final calculated results.  Also, it is clear that the given geometry sets physical limits for the minimum angle of incidence, $\theta_{\min}$,
given by $\theta_{\min}=a/L$, where $L$ is the distance between the observation and incident points.  
Thus in what follows, $\theta\rightarrow0$ implies $\theta\rightarrow\theta_{\min}$.

At grazing incidences, both the scattered waves and excited surface waves are interconnected.  To incorporate both effects, it is convenient to define 
the $\theta$-dependent total $Q_\theta$ factor,
\begin{equation}
\label{Qdef}
Q_\theta \equiv Q^s_\rho \sin\theta + Q^s_z \cos\theta,
\end{equation}
where
\begin{equation}
\label{Qs}
\begin{split}
Q^s_\rho &= \frac{R}{2\pi aI_0} \int_0^{2\pi} S^s_\rho(R,\phi) d\phi,  \\
Q^s_z &= \frac{1}{\pi (R^2-a^2) I_0} \int_a^R\int_0^{2\pi} S^s_z(\rho,\phi)\rho d\rho\,d\phi.
\end{split}
\end{equation}
Here $I_0=\frac{1}{2}{\cal Y}_0|E_0|^2$ is the input intensity (${\cal Y}_0$ is the vacuum admittance), and $R$ is the radius of the integration circle around the rod.  
The Poynting vector components $S^s_\rho$ and $S^s_z$ are along the $\hat\rho$ and $\hat z$ directions respectively, and are 
calculated from the scattered waves.  The $Q^s_\rho$ describes resonant scattering and the $Q^s_z$ describes guided surface-wave 
excitations.  The $Q_\theta$ is a useful metric to characterize the resonant scattering and the efficiency of energy channeling through the 
nanowire.  The $Q_\theta$ appropriatly reduces to the traditional $Q$ when $\theta=90^{\rm o}$ (normal incidence).  Our studies have shown 
that $Q^s_\rho$ is independent of $R$ for all points outside of the cylinder, as is the case for the traditional $Q$. 
To further elucidate the power flow in the $z$-direction, we define the effective flow area $A_e$:
\begin{equation}
\label{EFA}
A_e\equiv \frac{\int_a^R\int_0^{2\pi} S_z(\rho,\phi)\rho d\rho\,d\phi}{\max\left\{\int_0^{2\pi}S_z(\rho,\phi)d\phi\right\}} \,,
\end{equation}
where $S_z$ is calculated from the total field (scattered plus incident) outside the rod.  
The effective flow area is a function of the incident wavelength and the grazing angle, and it is useful in locating
resonant peaks.

\begin{figure}[htb]
\centering\includegraphics[width=.5\textwidth]{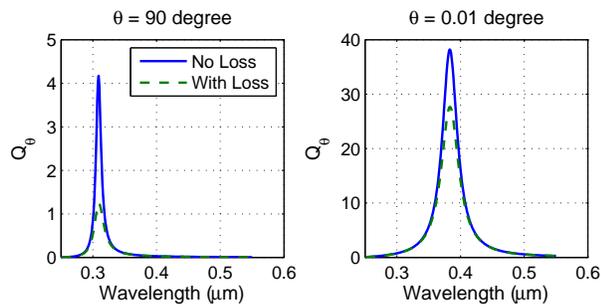}
\vskip-1.2in
\caption{Comparison of surface plasmon and cyclic Sommerfeld resonances:  Left: TE mode at normal incidence 
and SPR excitation at 309~nm.  Right: TE mode incident at the grazing angle of $0.01^{\rm o}$ 
and the corresponding resonance at about 380~nm.  Solid Blue: without material loss.  The $Q$ of the grazing resonance is about 10 times higher than that of SPR.  
Dashed Green: with material loss.  The $Q$ of the grazing resonance is then about 23 times higher than that of SPR.}
\label{GenPowTEplt}
\end{figure}

\begin{figure}[htb]
\centering\includegraphics[width=.5\textwidth]{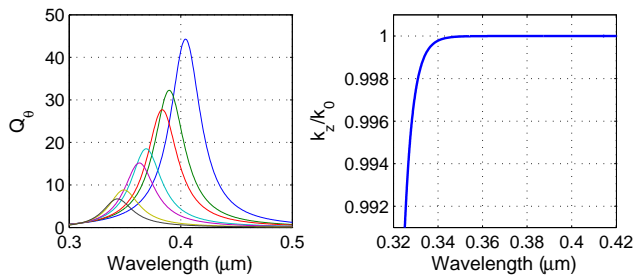}
\vskip-1.3in
\caption{Cyclic Sommerfeld resonances and Waveguide modes:  At the low grazing angles, a family of guided cyclic periodical ($n=1$)
surface waves can be excited.  Left plot: the peaks from low to high correspond to $\theta=1^{\rm o}$, $\theta=0.5^{\rm o}$, 
$\theta=0.1^{\rm o}$, $\theta=0.05^{\rm o}$, $\theta=0.01^{\rm o}$, $\theta=0.005^{\rm o}$, and $\theta=0.001^{\rm o}$.
The $Q$ factor and the wavelength of grazing resonances increases continuously as the grazing angle approaches the minimum angle.  Right plot: 
Calculated waveguide modes of the $n=1$ azimuthal mode at the low grazing angles for the silver nanorod.  These modes 
are related to the first-order grazing resonances shown on the left plot.  No cutoff frequency exists for the $n=1$ mode.}
\label{GrzPowVplt}
\end{figure}

The permittivity of the silver nanorods was obtained from a polynomial fit to experimental data\cite{Treacy},
appropriate for the visible spectrum of interest.  For the results below, 
we take $R=3a$ with $a=15~{\rm nm}$, and all field quantities are normalized with respect to the input field.
For a silver nanorod, the SPR at $\lambda\approx 309$~nm can be excited by a TE wave at normal incidence.  
The frequency of the SPR is insensitive to the rod diameter\cite{Fredkin}, however the $Q$ factor is 
sensitive to material loss.  At low grazing angles, this picture changes dramatically:
a series of non-plasmonic resonances emerge.  These resonances have a much higher $Q$ and broader bandwidth than the SPRs, 
as shown in Figs.~\ref{GenPowTEplt} and the left plot in Fig.~\ref{GrzPowVplt}.  Moreover, these grazing resonances are much 
less susceptible to material loss than the usual SPR (Fig.~\ref{GenPowTEplt}).  As the incident angle transitions from normal to grazing, the SPR continuously transforms into new resonant modes with a
corresponding redshift of the resonant frequencies.  This result is consistent with the calculated waveguide modes for the nanorod, shown in the right plot of Fig.~\ref{GrzPowVplt}.  As $k_z/k_0$ tends to unity, small changes in the grazing angle result in a significant shift in the resonant wavelength.  This explains the features found in the left plot of Fig.~\ref{GrzPowVplt}, where the resonant peaks continuously shift to longer wavelengths without cutoff when approaching the minimum angle.  Due to the material dispersion, the grazing resonances cannot in general be trivially scaled to other spectral regimes.  The Q factor for dielectric nanorods is about half of that for metallic rods of the same diameter\cite{Feng2}.
%%%%%%%%%%%%%%%%%%%%%%%% Figure 1 %%%%%%%%%%%%%%%%%%%%%%%%%%%%%%
%%%%%%%%%%%%%%%%%%%%%%%% Figure 2 %%%%%%%%%%%%%%%%%%%%%%%%%%%%%%
%%%%%%%%%%%%%%%%%%%%%%%% Figure 3 %%%%%%%%%%%%%%%%%%%%%%%%%%%%%%
\begin{figure}[htb]
\centering\includegraphics[width=.45\textwidth]{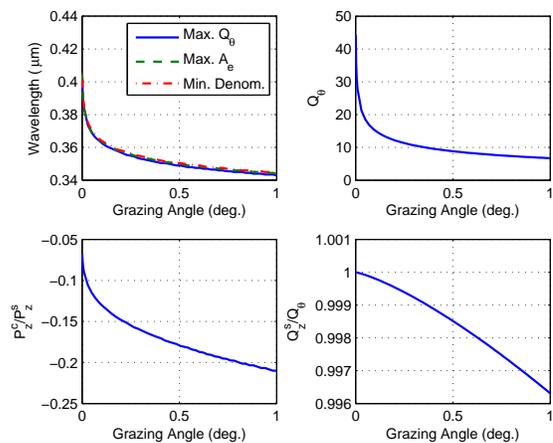}
\vskip-.1in
\caption{The $Q$ factors and power flows at near-zero grazing angles:  Upper-left: Wavelength of CSR vs. grazing angle calculated from the maximum $Q_\theta$ (Solid Blue), from the maximum $A_e$ (Dashed Green), and from the minimum of the $n=1$ denominator (Dashed-Dot Red).  The three curves coincide with each other.  Upper-right: Total $Q_\theta$ factor vs. grazing angle.  Lower-left: Ratio of the power flow in the z-direction inside ($P^c_z$) and outside ($P^s_z$) the rod.  Negative value is the result of the backward propagation inside the rod.  Lower-right: Ratio of the $Q^s_z$ and the total $Q_\theta$ factor.}
\label{GrzPowV10}
\end{figure}

The grazing resonances are correlated with the minimum of the denominator corresponding to the 
$n=1$ expansion coefficient of the scattered field, as shown by the red dashed-dot curve in the upper-left panel of Fig.~\ref{GrzPowV10}.  These coefficients represent the $n=1$ cyclic periodical surface waves.  Thus, the grazing resonances result from the excitation of first-order guided waves, which travel with a spiraling trajectory and constructive interference significantly enhances the Q factor of resonant scatterings.  It is interesting to examine the relationship between the grazing resonances and the resonant peaks extracted from $A_e$.  The upper-left panel of Fig.~\ref{GrzPowV10} demonstrates consistency among the three curves identifying the resonant wavelengths.  The enhancement of the $Q$ factor at near-zero grazing angle is due to guided asymmetric surface waves compounded with a coherent superposition of scattering states, which reinforces the resonant effect.  As the angle becomes smaller, $Q^s_z$, which describes the resonant guiding, increases and becomes dominant in the total $Q_\theta$ factor, as shown in the lower-right panel of Fig.~\ref{GrzPowV10}.  Another interesting phenomenon is that while the excited surface wave travels forward outside the nanorod, the power propagates backward inside the rod due to the boundary conditions and negative permittivity of silver.  The lower-left panel of Fig.~\ref{GrzPowV10} shows the ratio of the power inside ($P^c_z$) and outside ($P^s_z$) the rod in the $z$-direction.  As the angle decreases, the scattering in the radial direction becomes weaker and wave guiding along the rod becomes dominant.  Since the power inside the rod arises from radial scattering, the ratio $|P^c_z/P^s_z|$ decreases as $\theta\rightarrow\theta_{\min}$.
%%%%%%%%%%%%%%%%%%%%%%%% Figure 4 %%%%%%%%%%%%%%%%%%%%%%%%%%%%%%
\begin{figure}[htb]
\centering\includegraphics[width=.46\textwidth]{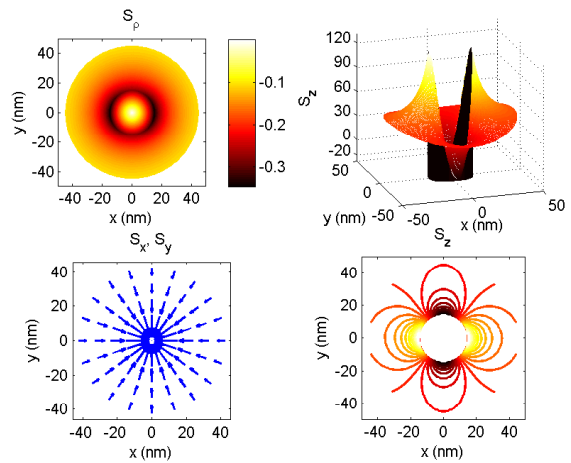}
\vskip-.1in
\caption{Radially symmetric scattering and backward propagation:
$\theta\approx0.001^{\rm o}$ at the resonance $\lambda=406$~nm.  Upper-left: Poynting vector $(S_\rho)$ in the $\rho$-direction.  The $S_\rho$ is nearly uniform in the x-y plane.  Lower-left: Vector plot of the Poynting vector $(S_x,S_y)$.  Power in the x-y plane flows radially towards the center of the rod.  Upper-right: The z-direction Poynting $(S_z)$ inside and outside the rod.  The $S_z$ is asymmetric and flows backward inside the rod.  Lower-right: Contour plot of the $S_z$ further showing the asymmetric distribution in the x-y plane.}
\label{PoynS17}
\end{figure}
%%%%%%%%%%%%%%%%%%%%%%%% Figure 5 %%%%%%%%%%%%%%%%%%%%%%%%%%%%%%
\begin{figure}[htb]
\centering\includegraphics[width=.48\textwidth]{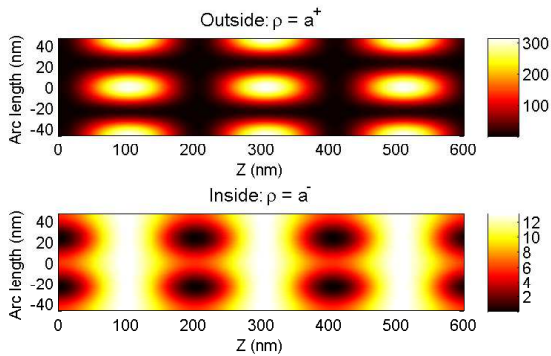}
\vskip-.8in
\caption{Intensity of the total electric field at the surface inside and outside the rod:  The intensity shows a 2D standing-wave pattern in the azimuthal and propagation directions.  Upper plot: Just outside the surface $\rho=a^+$.  Lower plot: Just inside the surface $\rho=a^-$.  Vertical axis: arc length along the circumference of the rod.  Horizontal axis: distance along the axis of the rod.}
\label{PoynS7}
\end{figure}
%%%%%%%%%%%%%%%%%%%%%%%% Figure 6 %%%%%%%%%%%%%%%%%%%%%%%%%%%%%%
\begin{figure}[htb]
\centering\includegraphics[width=.56\textwidth]{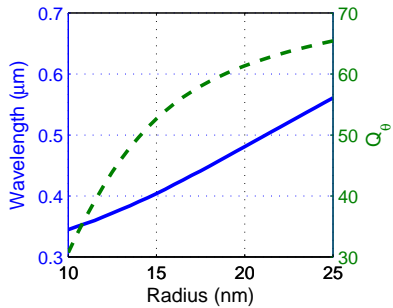}
\vskip-1.4in
\caption{Influence of the radius on grazing resonances:  $\theta=0.001^{\rm o}$.  Left y-axis for the solid-blue curve.  Right y-axis for the dashed-green curve.  For simplicity, the integral in $Q_\theta$ was performed at the surface $\rho=a^+$ over the azimuthal direction only.}
\label{GrzPowV9}
\end{figure}

To better visualize the spatial characteristics of the resonant fields, we show in Fig.~\ref{PoynS17} the Poynting vector from several perspectives at near-zero grazing incidence.  As shown in the left panel, the power ($S_\rho, S_x, S_y$) in the x-y plane is toward the center of the rod and has nearly perfect circular symmetry.  The degree of symmetry depends on how small the angle is.  However, once the energy penetrates the wire, it travels asymmetrically backward along the wire, as shown in the right panel.  Figure~\ref{PoynS7} shows the intensity of the total electric field close to the surface $\rho=a$ inside and outside the nanorod.  It reveals a two-dimensional standing wave pattern in the azimuthal and propagation directions, as the result of phase-matched spiraling propagation along the nanorod.  The geometric influence on grazing resonances is shown in Fig.~\ref{GrzPowV9}.  Both the resonant wavelength and the total $Q_\theta$ increase with increasing the radius of the nanorod.

In summary, we have investigated resonant scatterings of nanowires at near-zero grazing incidences through a newly developed grazing solution.  We also interpreted grazing resonances in terms of cyclic Sommerfeld waves and pointed out the non-plasmonic nature of grazing resonances.  This result enriches ones fundamental understanding of scattering on the nanoscale and its relevance to many areas within nanotechnology.  Since grazing resonances are associated with the natural modes of a nanorod, they may also be excited by other means.  The merit of high $Q$, broadband, and low loss may render this type of field enhancement as an attractive alternative mechanism for enhanced-field applications in the nano-regime.

The authors gratefully acknowledge the sponsorship of program manager Dr. Mark Spector and NAVAIR's ILIR program from ONR.

\end{document}